\documentclass[final]{svjour3}
\usepackage{graphicx}
\usepackage{rotating}
\usepackage{amssymb}
\usepackage{mathptmx}
\usepackage[numbers]{natbib}
\makeatletter
\journalname{Journal of Low Temperature Physics}

\bibpunct{}{}{,}{s}{}{,}

\usepackage{url}

\begin{document}

\newcommand{\hdblarrow}{H\makebox[0.9ex][l]{$\downdownarrows$}-}
\title{The Cryogenic AntiCoincidence detector for ATHENA X-IFU: improvement of the test setup towards the Demonstration Model}

\author{
M.~D'Andrea\textsuperscript{1,2} \and C.~Macculi\textsuperscript{1} \and A.~Argan\textsuperscript{1} \and S.~Lotti\textsuperscript{1} \and G.~Minervini\textsuperscript{1} \and L.~Piro\textsuperscript{1} \and M.~Biasotti\textsuperscript{3} \and D.~Corsini\textsuperscript{3} \and F.~Gatti\textsuperscript{3} \and G.~Torrioli\textsuperscript{4} \and A.~Volpe\textsuperscript{5}
}

\institute{ 
\email{matteo.dandrea@iaps.inaf.it} \\
\textsuperscript{1}INAF/IAPS Roma, Rome, Italy\\
\textsuperscript{2}Dept. of Physics, Univ. of Rome "Tor Vergata'', Rome, Italy\\
\textsuperscript{3}Dept. of Physics, University of Genova, Genoa, Italy\\
\textsuperscript{4}CNR/IFN Roma, Rome, Italy\\
\textsuperscript{5}ASI, Rome, Italy
}

\titlerunning{The CryoAC for ATHENA X-IFU: improvement of the test setup towards the DM}      
\authorrunning{M. D'Andrea et al.}

\maketitle

\begin{abstract}

The ATHENA X-IFU development program foresees to build and characterize an instrument Demonstration Model (DM), in order to probe the system critical technologies before the mission adoption. In this respect, we are now developing the DM of the X-IFU Cryogenic Anticoincidence Detector (CryoAC), which will be delivered to the Focal Plane Assembly (FPA) development team for the integration with the TES array. Before the delivery, we will characterize and test the CryoAC DM in our CryoLab at INAF/IAPS.
\newline In this paper we report the main results of the activities performed to improve our cryogenic test setup, making it suitable for the DM integration. These activities mainly consist in the development of a mechanichal setup and a cryogenic magnetic shielding system, whose effectiveness has been assessed by FEM simulations and a measurement at warm. The preliminary performance test has been performed by means of the last CryoAC single pixel prototype, the AC-S8 pre-DM sample.


\end{abstract}

\section{Introduction}


ATHENA\cite{athena} (Advanced Telescope for High ENergy Astrophysics) is an ESA large-class mission selected to be launched in 2030, consisting of a large X-ray telescope and two complementary focal plane instruments: the Wide Field Imager (WFI\cite{WFI}) and the X-ray Integral Field Unit (X-IFU\cite{XIFU}). The X-IFU is a cryogenic imaging spectrometer with unprecedented spectral capabilities ($\Delta$E $<$ 2.5 eV at 6 keV), based on a large array of about 4000 TES microcalorimeter. To enable the observation of faint and diffuse sources, thus meeting the scientific requirements of the mission, it is necessary to reduce the particle background in the array by adopting the TES-based Cryogenic Anticoincidence detector (CryoAC). Several CryoAC prototypes have been developed and tested by our consortium so far\cite{LTD15}$^,$\cite{LTD16}, in order to better understand the involved physics and define the final pixel design. 

We are now developing the CryoAC DM, which will be a single-pixel prototype able to probe the detector critical technologies (i.e. the threshold energy of 20 keV and the operation with a 50 mK thermal bath ). It will be preliminary characterized and tested in our cryogenic setup at INAF/IAPS, and then delivered to the FPA development team in order to be integrated and tested with the TES array. This will be the first compatibility test for the two detectors, representing a milestone on the path towards the X-IFU development.

The aim of this paper is to show the improvement of the cryogenic setup that we have achieved in preparation to the DM test activity. We will report the integration in our cryostat of a new magnetic shielding system, assessing its effectiveness by means of FEM simulation and a measurement at warm (Sect.~2). Then we will show the preliminary performance test carried out with the last CryoAC single-pixel prototype (Sect.~3), focusing on the effect of the Pulse Tube operations on the detector noise spectra (Sect.~4). Finally we will present the conclusions (Sect.~5).



\section{The new cryogenic magnetic shielding system}

It is generally known that the performances of TES microcalorimeters are strongly influenced by external magnetic fields down to the $\mu$T level\cite{ishisaki}. An appropriate magnetic shielding is therefore required to properly operate these detectors. In this respect, we have upgraded our cryostat inserting a cryogenic shielding system at the 2.5 K stage.
The cryostat is a commercial ADR pre-cooled by a Pulse Tube Refrigerator (PTR). The ADR superconducting magnet (shielded by the manufacturer) generates a maximum of $10^{-5}$ T in the experimental volume\cite{vericold}, so the dominant field to be shielded is the geomagnetic one ($\sim$ $ 0.5 \cdot 10^{-4}$ T). 

\begin{figure}[htbp]
\begin{center}
\includegraphics[width=0.8\linewidth]{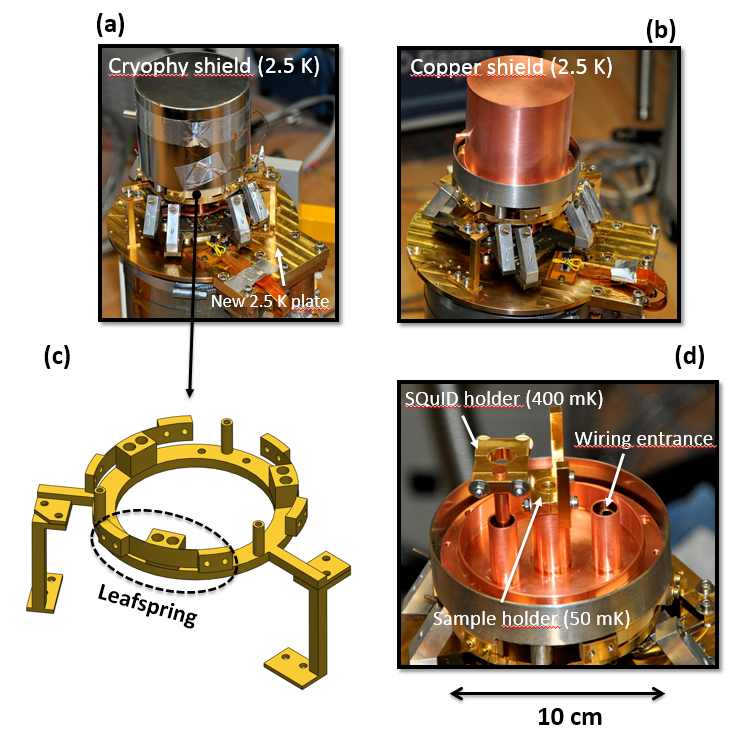}
\caption{New cryostat setup with the cryogenic magnetic shielding system mounted on the 2.5K plate. (a) The external Cryophy shield. (b) The inner copper shield. (c) CAD model of one mechanical structure supporting the shields. (d) The shielded sample area. (Color figure online)}
\end{center}
\label{fig1}
\end{figure}

We have developed a shielding system constituted by three main parts (Fig.~1). The first part is a ferromagnetic cylindrical shield made of Cryophy\cite{cryophy} (Fig.~1a), a nickel-iron soft alloy with high magnetic permeability ($\mu_r \sim 10^5$ at cold), suitable for static and low-frequencies magnetic shielding at cryogenic temperature. The shielding factor S of this shield (i.e. the ratio of the internal magnetic field to the external one) can be roughly estimated as\cite{shielding}:
\begin{equation}
S \sim \left( \frac{D}{\mu_r \cdot d} \right)^{-1} \sim 500
\end{equation}
where D = 100 mm and d = 1 mm are the diameter and the thickness of the shield, and $\mu_r = 0.5\cdot 10^5$ @ 4K is the initial permeability measured by the manufacturer on a Cryophy sample ring. We therefore expect a residual static magnetic field inside the shield of about $10^{-7}$ T.

The second element in the system is an OFHC copper shield (Fig.~1b) that, if necessary, will be lead plated to operate as a superconducting shield. It is placed inside the Cryophy in order to find an appropriate low field environment during the superconductive phase transition, avoiding flux trapping.

Third, we have developed a gold-plated copper structure (Fig.~1c) to mimimize the mechanical stress on the Cryophy shield during the cooling phase, balancing the thermal contractions. This is important for an efficient shielding, as the magnetic permeability of the material rapidly decreases with the stress level\cite{sron}. The core of this structure are three leafspring-like elements with flexible arms (300 $\mu$m thick). 

The shields present three openings to provide the access into the experimental area (Fig.~1d). Two of these openings are used to host two rods connected to the first ($\sim$ 400 mK) and the second ($\sim$ 50 mK) ADR stages, where the SQUID holder and the sample holder structures have been anchored. The third aperture is instead designed for the wiring entrance (i.e. for TES bias, SQUID read-out and thermometry).

Finally, we have also upgraded the 2.5 K plate of the cryostat, moving from Aluminum to gold-plated Copper, in order to minimize thermal gradients within the plate and do not affect the Pulse Tube performances.

\subsection{Magnetic field modeling and measurements}

To model the magnetic field attenuation inside the new shielding system, we have performed a 3D Finite Element Method (FEM) simulation using the COMSOL Multiphysics \textregistered \cite{comsol} software package. The simulation consists in a stationary study including the \textit{Magnetic Fields, No Currents (mfnc)} physics interface of the \textit{AC/DC module} \cite{comsol2}.

The system geometry has been implemented importing in the software the CAD drawings of the shields, and the mesh has been refined until no significant variation in the simulation results was observed. The shields have been placed in a spatially uniform background magnetic field $B_{bkg}$ of strenght $10^{-4}$ T, alternately oriented parallel or perpendicular to their main axis. The cryophy relative permeability has been defined as a constant value: $\mu_r = 51000$, equal to the initial permeability measured at 4 K by the manufacter on a Cryophy sample ring subjected to the same annealing process of the shield. Being the annealing performed at zero magnetic field, the magnetization properties of the material are described by the initial magnetization curve\cite{shielding2}, and so this value is the most representative for the relative permeabilty of the shield in a low magnetic field. 

The main results of the simulation are shown in Fig.~2, where the color scale indicates the shielding factor ($ S = (|B|/B_{bkg})^{-1} $) evaluated on a cut plane crossing the main shields axis. The magnetic flux density lines are overplotted in light blue, showing the field distortion induced by the Cryophy layer (note that for visual purpose the line density is not proportional to the field strenght).

\begin{figure}[htbp]
\begin{center}
\includegraphics[width=0.99\linewidth]{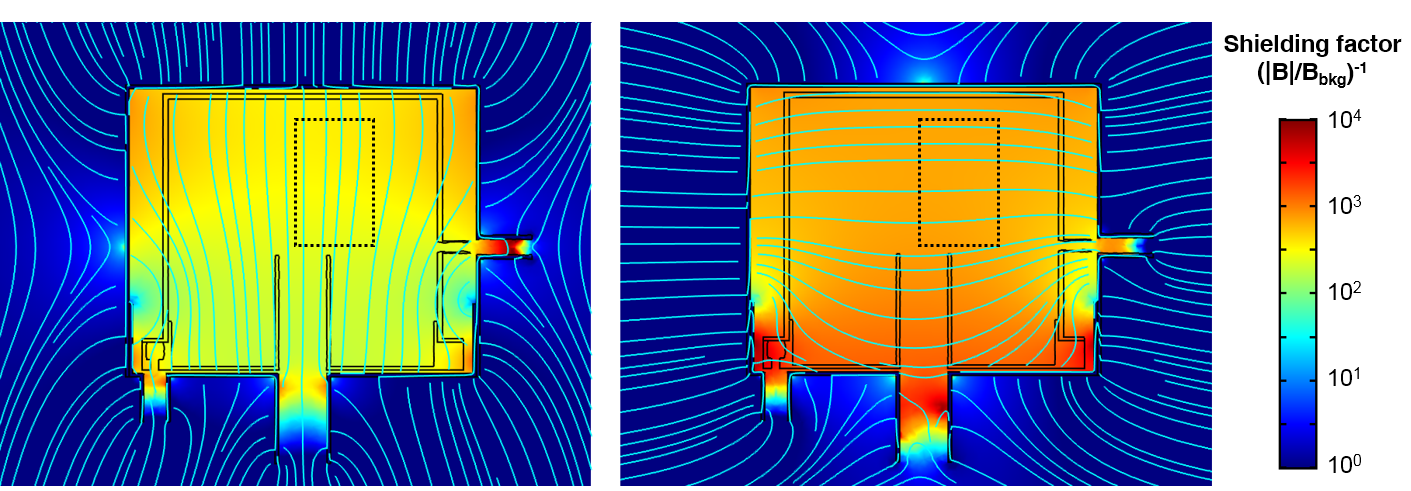}
\caption{Attenuation of the magnetic field inside the new shielding system evaluated by a FEM simulation. The color scale indicates the shielding factor on a cut plane crossing the main shields axis. The magnetic flux density lines are overplotted in light blue, and the detector zone is highlighed by the black dotted line. \textit{Left:} Background magnetic field in the shield axial direction. \textit{Right:} Background magnetic field in the shield transverse direction. (Color figure online)}
\end{center}
\label{fig1b}
\end{figure}

In Tab.~1 are reported the minimum and the maximum value of $S$ evaluated inside the shields and in the detector zone (black dotted region in Fig.~2), for both the axial and transverse directions of the background field. Note that the results of the simulation agree within a factor $\sim$ 3 with the rough estimation in Eq.~(1).

\begin{table}[htbp]
\centering
\caption{Shielding factors evaluated by FEM simulation}
\label{tab:DMreq}
\begin{tabular}{cccccc}
\hline\noalign{\smallskip}
Background field  & \multicolumn{2}{c}{Inside the shields} & \multicolumn{2}{c}{In the detector zone} \\
direction & $S_{MIN}$ & $S_{MAX}$ & $S_{MIN}$ & $S_{MAX}$ \\
\noalign{\smallskip}\hline\noalign{\smallskip}
Axial & 174 & 501 & 251 & 398 \\
Transverse & 501 & 1573 & 631 & 794 \\ 
\noalign{\smallskip}\hline
\end{tabular}
\end{table}

Finally, to preliminary check the simulation results we have performed a measurement of the axial shielding factor in the detector zone by a flux gate magnetometer: a HP Clip on DC Milliammeter 428B equipped with a magnetometer probe. The measurement has been performed at room temperature, fixing the position of the probe and measuring the axial ambient magnetic field with and without the shields around it. The procedure has been then repeated orienting the probe in the opposite direction, in order to correct the systematics due to the instrument zero-setting calibration. We remark that the Cryophy initial permeability does not show any significant degradation from room to liquid helium temperature\cite{cryophy2}, so we expect this measurement to be representative of the shielding effectiveness at the working temperature. We obtained:

\begin{equation}
S_{\textit{\scriptsize axial, measured at detector zone}} = 315 \pm 15
\end{equation}

\smallskip

\noindent The measured value is well consistent with the simulation results reported in Tab.~1.\newline
\newline
Given the earth magnetic field ($\sim$ $ 0.5 \cdot 10^{-4}$ T), we therefore expect in the detector zone a residual static magnetic field $< 2 \cdot 10^{-7}$ T.

\section{ADR and sample holder thermal performance, and preliminary test of AC-S8}


We have performed a preliminary test of the new cryogenic setup with the last CryoAC single pixel prototype, namely AC-S8. The sample is based on a large area (1 cm$^2$) Si absorber sensed by 65 Ir TES coupled to a network of Al finger (for more details see the AC-S8 reference paper in this issue\cite{acs8}). The detector has been coupled to a commercial SQUID array (Magnicon C6X216FB) and operated with a FLL electronics (Magnicon XXF-1). The test setup is shown in Fig.~3. 

\begin{figure}[htbp]
\begin{center}
\includegraphics[width=0.7\linewidth]{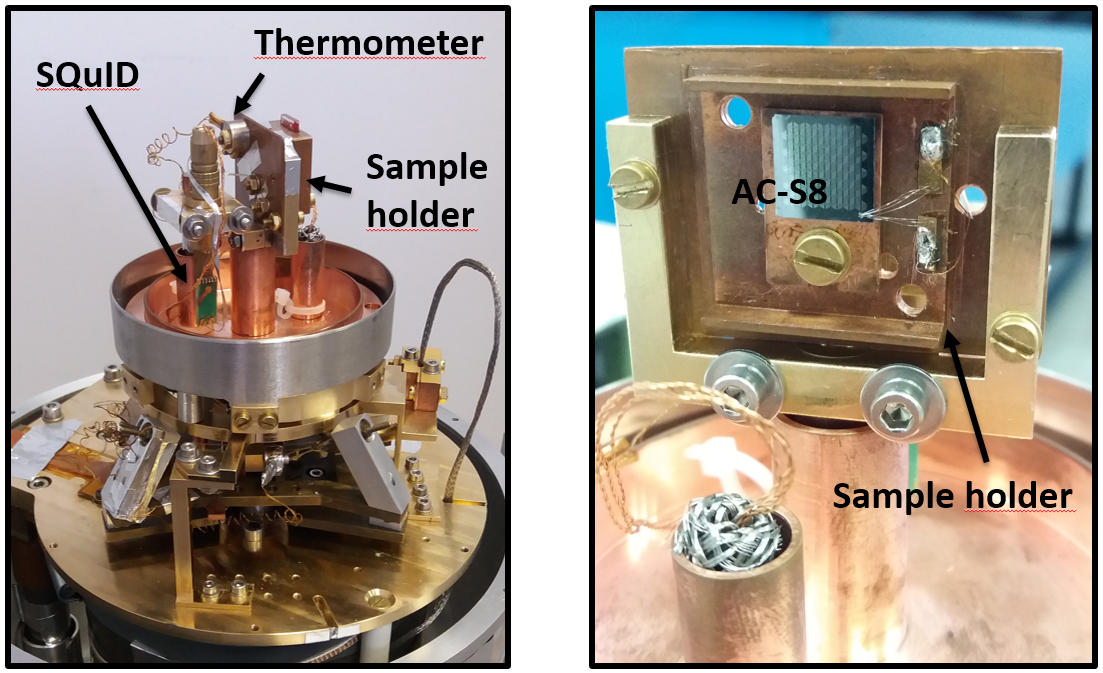}
\caption{\textit{Left}: Setup for the preliminary performance test of the new shielding system. \textit{Right:} Detail of the AC-S8 prototype mounted on the sample holder (before being covered by a copper sheet). (Color figure online)}
\end{center}
\label{fig2}
\end{figure}

First, we have performed a cooling test. 
The result is shown in Fig.~4, where are reported the temperature profiles of the cryostat stages acquired during an ADR cycle. Note that the sample stage (red curve in the plot) is able to reach a base temperature of 36 mK, remaining below 50 mK for about 10 hours. The absolute temperature accuracy is due to thermometer calibration uncertainty, which is evaluated to be $\sim$ 2 mK at T = 50 mK.  The cryostat and sample holder cooling performances are therefore fully suitable for our planned test activities, which foresee to operate the CryoAC DM with the thermal bath regulated at 50 mK, for which we expect an hold time higher than 10 hours measured at open loop.

\begin{figure}[htbp]
\begin{center}
\includegraphics[width=0.6\linewidth]{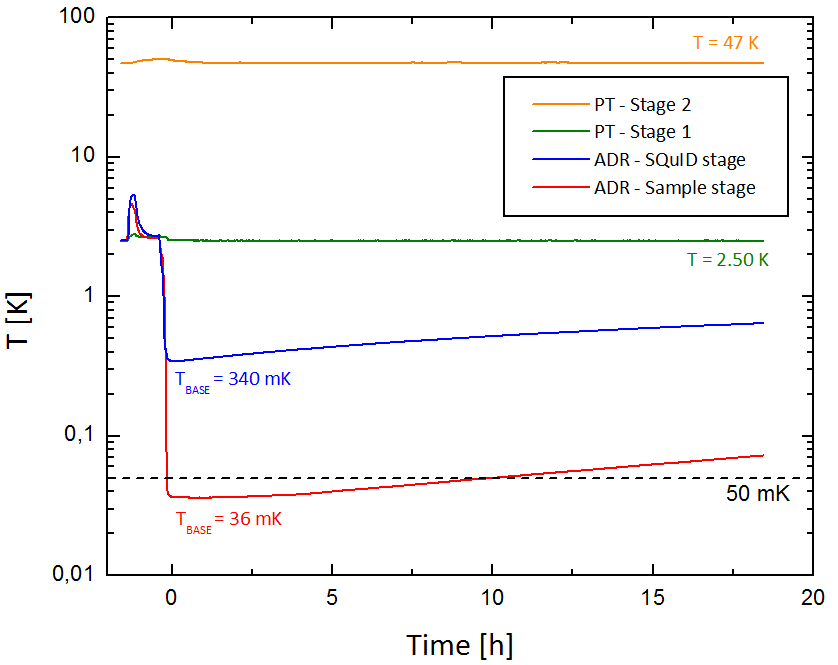}
\caption{Temperature profiles of the cryostat stages measured during an ADR cycle. (Color figure online)}
\end{center}
\label{cooling}
\end{figure}

We have then tested how the presence of the new shields could affect the TES superconductive transition. We have performed three consecutive measurements of the R vs T curve: with the shields, without the shields and again with the shields (to replicate the initial conditions). The results are reported in Fig.~5, which shows that with the shields the critical temperature of the sample results about 20 mK higher (from 99 mK to 120 mK). We interpret this as an evidence of the reduction of the static magnetic field perpendicular to the detector surface (which strongly influence the superconductive transition\cite{ishisaki}). We remark that the three measurements have been performed with the same method (SQUID readout, modulation technique with a low amplitude sine wave at 22 Hz) and that the magnetic shielding does not influence the thermal radiation on the AC-S8 sample (which is totally covered by the holder).
\newline

\begin{figure}[htbp]
\begin{center}
\includegraphics[width=0.59\linewidth]{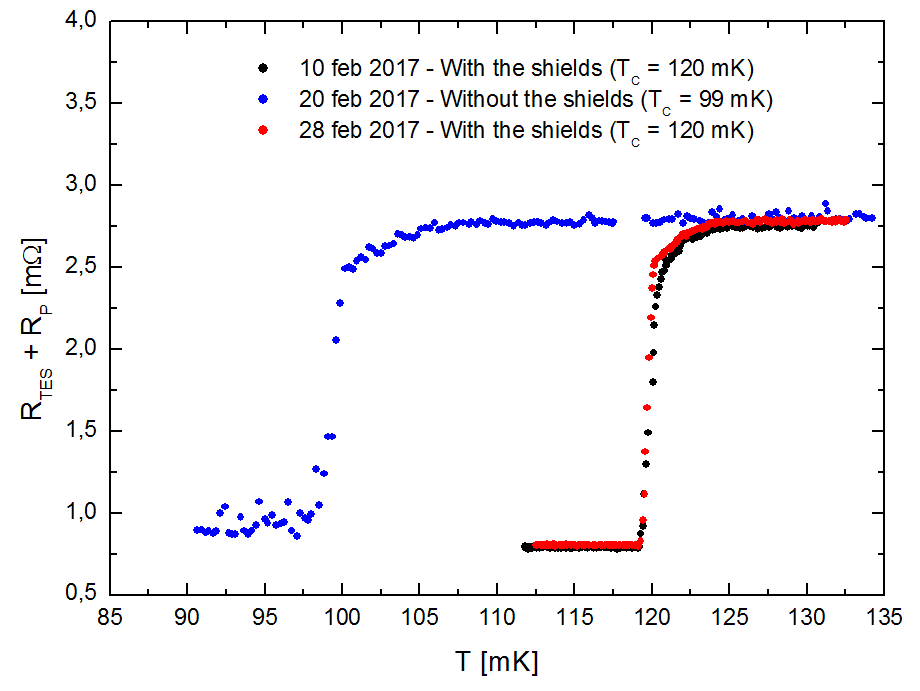}
\caption{Comparison between the AC-S8 transition curves measured with and without the shields. The critical temperature are evaluated at the half of the transition. (Color figure online)}
\end{center}
\label{rts}
\end{figure}

Lastly, we have performed a noise measurement in order to roughly quantify the effectiveness of the new shielding system also for non-static magnetic fields. In Fig.~6 are reported the noise spectra measured with the detector in the normal state (T = 130 mK), with and without the shields. Both the spectra have been acquired switching off the PTR compressor, in order to minimize the effect of the induced vibrations (in the next section we will focus on this point). The expected Johnson noise, filtered by the L/R of the TES circuit, is overplotted for comparison. The plots show that the introduction of the new shielding system lead to a reduction of the spectral noise lines, especially in the low-frequency region (as expected for Cryophy, which acts up to some hundred Hz). To quantify this improvement of the setup, we have integrated the noise spectra to obtain the equivalent current noise I$_{noise}$ in the 10 Hz - 100 KHz band. The results are reported in the plot, showing that the shields are able to reduce I$_{noise}$ by a factor $\sim$ 2 (from 7 nA to 3 nA RMS), reaching the theoretically expected Johnson value.

\begin{figure}[htbp]
\begin{center}
\includegraphics[width=0.6\linewidth]{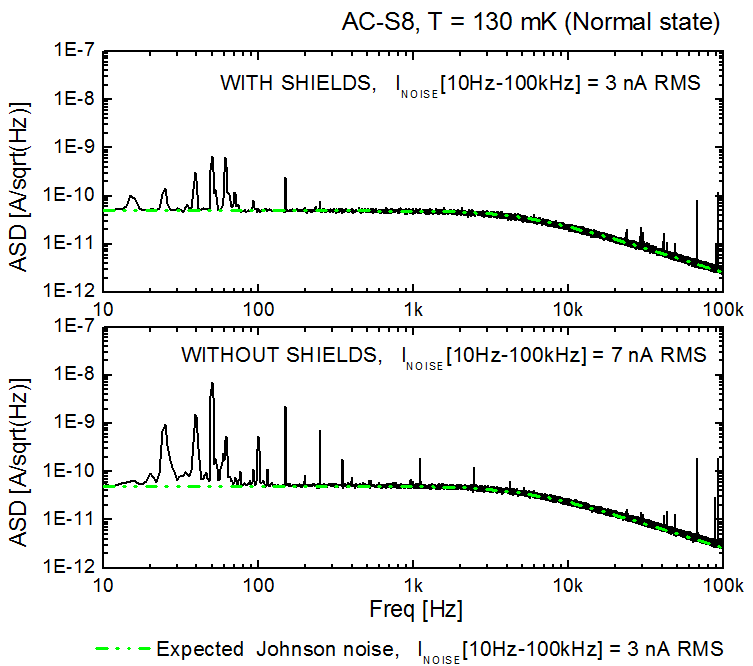}
\caption{Current noise Amplitude Spectral Density measured for AC-S8 in the normal state and with PTR turned OFF, with the shields (Top) and without them (Bottom). The dashed green line represents the expected Johnson noise.(Color figure online) }
\end{center}
\label{fig4}
\end{figure}

\section{The effect of the Pulse Tube operation}

\begin{figure}[htbp]
\begin{center}
\includegraphics[width=0.6\linewidth]{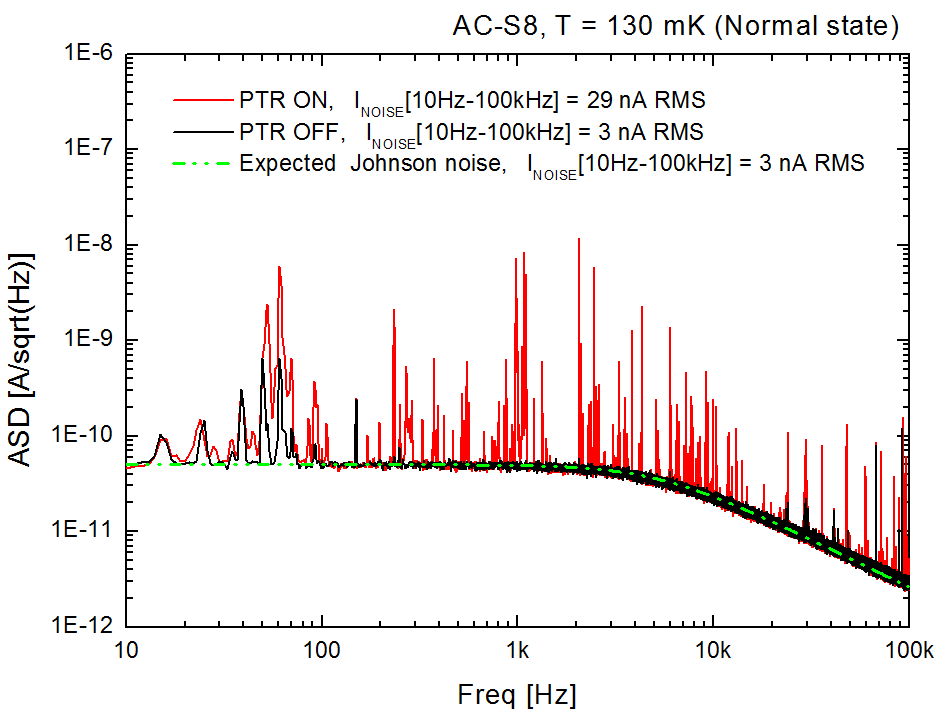}
\caption{Current noise Amplitude Spectral Density measured for AC-S8 in the normal state and both the shields mounted, with the PTR compressor switched ON (red curve) and turned OFF (black curve). (Color figure online)}
\end{center}
\label{fig5}
\end{figure}

Despite the good results obtained with the introduction of the new magnetic shielding system, we have found that the PTR operation has a dominant impact on the detector noise spectrum. In Fig.~7 are reported the spectra acquired with AC-S8 in the normal state (T = 130 mK) and the PTR compressor switched on (red line) and then turned off (black line). In both the cases the shields were mounted. The noise spectrum degradation is evident, with an increment of the equivalent current noise of one order of magnitude (from 3 nA to 29 nA RMS). This is probably due to an electromagnetic component and to the mechanical vibrations induced by the PTR on the cold stages of the cryostat, which cause microphonic disturbance in the TES system. This kind of effects are reported in literature, with an influence on the noise spectrum up to tenths of kHz and characteristics lines around 1 KHz\cite{ptr}, as in our case. 

We remark that we are, however,  able to switch off the PTR and still operate the detector for a while, typically about 15 min with the bath temperature at 50 mK. Therefore, this problem does not prevent us to perform the planned DM characterization activities.


%
%

\section{Conclusions}

We have here presented the main improvements in the cryogenic setup that we achieved in preparation to the CryoAC DM integration activity. 

We have developed a cryogenic magnetic shielding system, and assesed by means of FEM simulation and a test at warm that it is able to provide a shielding factor $S > 250$ in the detector region. Given the earth magnetic field ($\sim$ $ 0.5 \cdot 10^{-4}$ T), we therefore expect in the detector zone a residual static magnetic field $< 2 \cdot 10^{-7}$ T. After the system integration, the cryostat and sample holder thermal performances are fully suitable for the planned DM activity (hold time at 50 mK higher than 10 hours). 

We have performed a preliminary performance test of the magnetic shielding by means of the CryoAC AC-S8 prototype. We have found that inside the new shields the sample shows a critical temperature $\sim$ 20 mK higher than the one measured without the shields (T$_C$ from 99 mK to 120 mK). This is an evidence of the static magnetic field reduction over the detector surface. As expected, the new shielding system has also led to a reduction of the spectral noise lines at low frequency (up to some hundred Hz), improving by a factor $\sim$ 2 the measured equivalent current noise. 

Finally, we have shown the dominant effect of the PTR operations on the noise spectra, concluding that we are able to properly operate the detector only switching off the PTR (T$_{BATH} = 50$ mK typically maintained for $\sim$15 min). In this regards, we report that we 
have recently installed a new dry Dilution Refrigerator, and we are now  moving the magnetic shielding system in the new cryostat. This will allow us to have more available cooling power (450 $\mu$W @ 100 mK) and could also operate for more time with the PTR switched off (several hours with the cold stage at 50 mK).

\begin{acknowledgements}
This work has been partially supported by ASI (Italian Space Agency) through the Contract no. 2015-046-R.0. The authors would like to thank Henk van Weers (SRON) for useful discussion about the magnetic shielding development.
\end{acknowledgements}

\end{document}